\documentstyle[12pt,aasms4]{article}

\slugcomment{Submitted to the Astronomical Journal}
\begin{document}

\title{Massive star clusters in ongoing galaxy interactions -- clues
to cluster formation
\footnote{Based on observations with the NASA/ESA {\it Hubble Space Telescope}
obtained at the Space Telescope Science Institute, which is operated
by the Association of Universities for Research in Astronomy, Inc.,
under NASA contract No. NAS5-26555.}}

\author{William C. Keel\altaffilmark{2}}
\affil{Department of Physics and Astronomy, University of Alabama, Box 870324,
Tuscaloosa, AL 35487}

\author{Kirk D. Borne}
\affil{Institute for Science and Technology, Raytheon (IST@R)
and NASA Goddard Space Flight Center, Greenbelt, MD 20771}

\altaffiltext{2}{Visiting Astronomer, WIYN Observatory, which is owned and 
operated by 
the WIYN Consortium, Inc., which consists of the University of Wisconsin, 
Indiana University, Yale University, and the National Optical Astronomy 
Observatory (NOAO), and at Kitt Peak National Observatory, NOAO. NOAO 
is operated for the National Science Foundation 
by the Association of Universities for Research in Astronomy (AURA), Inc.}

\begin{abstract}
We present HST WFPC2 observations, supplemented by ground-based H$\alpha$ data,
of the star-cluster populations in two pairs
of interacting galaxies, selected for being in very different kinds of
encounters seen at different stages. Dynamical information and
$n$-body simulations provide the details of encounter geometry, mass ratio,m
and timing. In NGC 5752/4, we are seeing a weak encounter well past
closest approach, after about $2.5 \times 10^8$ years. The large spiral 
NGC 5754 has a normal population
of disk clusters, while the fainter companion NGC 5752 exhibits a
rich population of luminous clusters with a flatter luminosity function.
The strong, ongoing encounter in NGC 6621/2, seen about $1.0 \times 10^8$
past closest approach between roughly equal-mass galaxies,
has produced an extensive
population of luminous clusters, particularly young and luminous
in a small region between the two nuclei. This region is dynamically
interesting, with such a strong perturbation in the velocity field that the
rotation curve reverses sign. From these results, in comparison with
other strongly interacting systems discussed in the literature, 
cluster formation requires a threshold level of perturbation, with
stage of the interaction a less important factor. The location of the most 
active star formation in NGC 6621/2 draws attention to a possible role for
the Toomre stability threshold in shaping star formation in interacting
galaxies. The rich cluster populations in NGC 5752 and 6621 show that
direct contact between gas-rich galaxy disks is not a requirement to
form luminous clusters, and that they can be triggered by processes
happening within a single galaxy disk (albeit triggered by external
perturbations).

\end{abstract}


\section{Introduction}

Evidence connecting galaxy interactions to powerful star-forming events
has accumulated from observations over a wide range of wavelengths,
from the radio continuum to X-rays. This enhancement in star formation
does not consist merely of exceptional numbers of normal H II
regions. Among the most striking results
from early HST observations of interacting and merging systems
was the prevalence
of luminous, presumably massive super star clusters (SSCs). While a
handful are sufficiently bright and isolated to have been picked out from
earlier ground-based data (e.g. Schweizer 1982 and Lutz 1991), it was the 
improved resolution of HST (for an early example, using pre-refurbishment
data, see Whitmore et al. 1993)
which revealed
them as an important population. In particular, their high luminosity
allows the possibility that these are in effect young globular clusters,
if they have a ``normal" initial-mass function continuing to low
stellar masses. This would fit with the idea that strongly interacting
and merging galaxies today can serve as a useful model for conditions
early in galaxy formation, when more isolated galaxies (such as ours)
formed similar clusters as well.

As reviewed by Whitmore (2002), much of the focus in relating SSCs to
galaxy interactions has been in following the cluster populations
during and after major mergers. These studies have shown a general
correlation between the ages of the SSCs, as derived from modelling their
broadband colors, and the age of the merger from dynamical considerations.
Objects seen close to the time of merger - such as infrared-luminous
galaxies - can have extensive populations of such clusters in their
inner few kiloparsecs. The benchmark for young cluster population has been the 
Antennae, NGC 4038/9 (Whitmore et al. 1999b). The remnant disks of both
galaxies contain rich populations of clusters as bright as
$M_V=-14$, with evidence for distinct star-forming events in various
parts of the system. The luminosity function of these clusters
is close to a power law with slope $\alpha=-2.1$, possibly more
precisely described as a broken power-law form. The luminosity function
of SSCs, and its possible evolution, are important in probing the
connection between these young, massive clusters and old globular
clusters.

Much of the interest in these cluster populations has centered on the
question of whether they are massive enough to be genuine globular
clusters, and how this fits with the merging hypothesis for the
formation of elliptical galaxies. This has driven many studies of
the SSC population to concentrate on merging systems of various
dynamical properties (forming a rough age sequence), and a few
galaxies seen very strongly interacting.
As reviewed by Whitmore (2002), there is a continuous color sequence
which follows a plausible fading line from the young clusters in the
Antennae to globular clusters in such ellipticals as M87. While this
favors the idea that these are similar populations seen at various
epochs, and thus that mergers form many elliptical galaxies and
important parts of their globular-cluster systems, the luminosity
function of SSCs in many systems is well described as a power law,
like that of ordinary open clusters in spiral galaxies, and quite
different from the approximately Gaussian distribution of the
old, ``genuine" globular clusters in both spirals and ellipticals. This 
makes the survival and evolution of the clusters over
Gyr timespans crucial in telling whether there really is a continuous
sequence all the way to old globular clusters.

In contrast to these consideration of the long-term evolution of these
luminous clusters, rather few galaxies in ``typical" interactions outside of
merging conditions have been
surveyed for SSCs, to assess a link between immediate tidal
events and star formation in such massive concentration. 
Such interacting pairs may offer a cleaner view of when and how clusters form,
with the dynamical history more easily recovered than in advanced mergers,
and different regions still connected to different tidal histories.
It would be especially helpful if we see very different star-formation
responses in two gas-rich galaxies in a pair, since the timing
of encounter is identical.
In an effort to specify what kinds of galaxy dynamics and timescales
are involved in the formation of SSCs, we have observed the cluster
populations in two interacting galaxy pairs in different extremes of
the interaction process, both well before merging takes place. Using 
morphology and $n$-body simulations,
we selected pairs that are early in a very strong tidal encounter
(NGC 6621/2), and that are at a late stage in a weak encounter
(NGC 5752/4, which is weak for the larger galaxy at least). One might 
expect some kind of threshold for formation
of these massive clusters, since they are not numerous in undisturbed
disks), and may require such special processes as collisions of
giant molecular clouds or assembly of massive clouds within a particular
time. Larsen (2002) finds that only a fraction $\sim 15$\% of luminous
``normal" spirals have any cluster as bright as $M_V=-12.0$,
and that very rich cluster populations are needed to expect such
examples for a typical power-law luminosity function.
Given this evidence that the cluster populations in interacting systems 
result from unusual processes, the systems for which we can 
infer ages and levels of perturbation might then offer clues
to where and when the clusters form.

\section{Observations}

\subsection{Selection and properties of galaxy pairs}

We selected candidate pairs starting with the samples observed by Keel et al. 
(1985) spectroscopically and in H$\alpha$ imaging, to select systems with active
star formation, so we would be able to ask how much of this
star formation occurs in SSCs. We sought pairs with morphologies
suggesting a single encounter of only two galaxies, so that
their histories could be inferred from comparison with the
$n$-body atlas results from Howard et al. (1993), and lower
redshifts for the highest sensitivity and linear resolution.
From these criteria, we settled on the pairs NGC 5752/4 and NGC 6621/2
to represent two extremes of the pair population. The pairs
also offer additional contrasts, in that NGC 5752/4 are of
very different size and luminosity, and NGC6621 is much more gas-rich
than NGC 6622.

NGC 5752/4 is part of the apparent quartet Arp 297. Despite the
interestingly close configuration of these galaxies, redshift
information breaks this grouping into two separately interacting pairs
differing in distance by a factor of two (Fig. 1). The
figure also shows the outer tidal structure which aids in reconstructing
the dynamical history of the pair, including a faint tail
running westward from NGC 5752. The redshift listed for NGC 5752 in NED
is somewhat suspicious, being exactly the same as listed for NGC 5754.
We have therefore re-examined the low-resolution
spectrophotometric data from Keel et al. (1985, where it is listed
as Arp 297C), from which we derive a 
heliocentric velocity $4325 \pm 85$ km s$^{-1}$ from H$\gamma$, H$\beta$, and 
[O III] emission. This agrees well with a newly
available archived spectrum from the CfA $z$-machine
\footnote{http://tdc-www.harvard.edu/cgi-bin/arc/uzcsearch} 
which gives $4441 \pm 24$ km
s$^{-1}$ from absorption-line cross-correlation. Adopting a value of 
${\rm H}_0 = 70$ km s$^{-1}$ Mpc$^{-1}$ gives a linear scale of 316
pc/arcsecond for this pair.

NGC 5754 exhibits a rich population of disk H II regions, and
NGC 5752 shows multiple, very bright H II regions closely packed around
its core (Kennicutt et al. 1987), satisfying our requirement for
substantial star formation.

NGC 6621/2 (Arp 81, VV 247, CPG 534) is much more strongly disturbed, 
kinematically as well as morphologically (Reshetnikov \& Sil'chenko 1990, 
Keel 1993, 1996). It is part of the original sequence of
candidate merging pairs presented by Toomre (1977).
The fainter galaxy in this case is an E/S0 system, whose redshift
is (like that of NGC 5752) in some question from published data. Gas from the
larger disk galaxy NGC 6621 is so widespread that emission lines
putatively seen from NGC 6622 are almost certainly from the
extended disk of NGC 6621, based both on the morphology 
and velocity continuity of the line emission. We therefore
measured an absorption-line value to help in interpreting the
history of this system (section 4.1). We adopt a linear scale of 
430 pc/arcsecond for this pair, based on a mean redshift
$cz = 6210$ km s$^{-1}$.

\subsection{HST Imaging}

Each galaxy pair was observed with WFPC2 in both $B$ (F450W) and $I$ (F814W)
passbands. As it happened, both were observed within a single 24-hour span, on 
14 and 15 March 1999. The filters were selected to provide a wide color 
baseline coupled with
high efficiency and rejection of strong emission lines. The elongated
pair NGC 6621/2 was placed in CCDs WF2 and WF3. For NGC 5752/4,
the high-surface-brightness companion NGC 5752 was placed on the
PC CCD, with NGC 5754 and its tidal arms spreading across all three
WF CCDs. This placement turned out to be fortunate because of the
crowding of the many luminous clusters found in NGC 5752.
Total exposure time in each filter was 2600 seconds, each
split into two individual exposures for cosmic-ray rejection.
Mosaics of the WFPC2 images are shown in Figs. 2 and 3. Data from the
individual CCDs have been rebinned to a common astrometric frame
for these figures, but cluster photometry was done on the original
data.

We followed the photometric procedures of Whitmore et al. (1999b)
as closely as the greater distance our targets would allow, to
be able to compare the cluster populations most directly with the
well-observed Antennae system NGC 4038/9. Cluster candidates were
identified using the IRAF 
\footnote{IRAF is distributed by the National Optical Astronomy Observatories,
which are operated by the Association of Universities for Research
in Astronomy, Inc., under cooperative agreement with the National
Science Foundation.} implementation of DAOFIND,
and measured using aperture photometry (radius 0.2"
and curve-of growth aperture correction of 0.70 magnitude) 
with a local background. For simplicity of analysis, we adopt
a uniform flux threshold for candidate clusters across each
galaxy, set by experiments with the number of false detections at
a given detection threshold, and removing a few spurious detections produced by
the greater Poisson noise near the center of NGC 5754. As in similar studies, 
cluster
colors are almost completely independent of aperture effects,
with only 2-7\% changes in the fraction of encircled energy
within the 0.2" aperture expected across our wavelength baseline.
We followed Holtzman et al. (1995) in assuming an aperture of
radius 0.5" to encompass 90\% of the energy. Between 0.2" and 0.5",
we find that the brightest clusters show a curve of growth 
similar to the point-source prediction (specifically, the
brightest ones in NGC 5754 give a correction 0.23 magnitude,
compared to a calculated value of 0.17 from the WFPC2 handbook).
Residual photometric errors due to geometric distortion within the
WFPC2 optics, which exist because the flat-field correction
is set up to properly correct surface brightness rather than point-source
intensity, should be less than 2\% for the regions spanned by these galaxies.
The role of foreground extinction should be quite modest for these
pairs; the Schlegel, Finkbeiner, \& Davis (1998) prescription gives
$A_B=0.20, A_I=0.09$ for NGC 6621/2 at $b=23^\circ$, and
$A_B=0.05, A_I = 0.02$ for NGC 5752/4 at $b= +63^\circ$. We fold these values
into the respective photometric zero points.
Charge-transfer effects are not very important even for the NGC 5752 data, 
taken in the PC CCD with relatively low background. We follow the prescription of
Dolphin (2000) in evaluating these, recognizing that the
complex brightness distribution makes these corrections
approximate. For the faintest clusters we measure, at the highest
instrumental $y$ coordinates, the correction is only 0.025 magnitude
in $B$, with a differential $B-I$ correction of $-0.006$ magnitude.
Using the prescription from Whitmore et al. (1999a) gives slightly
higher values, 0.04 magnitude in $B$, but these effects are still swallowed
up in the other photometric errors for these clusters.

Near the bright nucleus of NGC 5754, some nonexistent objects
appeared in the original source list simply due to the greater
Poisson noise and amplitude of local image structure. We rejected
these based on a median-windowed image, which removed most of the
effects of a strong brightness gradient on the object-finding
routine.

At the distance of either galaxy pair, genuine SSCs should be 
at most marginally resolved by WFPC2. However, since some of the detections in 
nearby systems such as NGC 4038/9
have dimensions of a few hundred parsecs, being perhaps better described
as superassociations, we did not require candidate clusters to
be completely stellar in structure for inclusion. This may also
include some blended sets of clusters, which can sometimes be
distinguished by image structure. At these distances and our
magnitude limit, confusion between candidate clusters and supergiant stars
will not be an issue, since B=26 at the distance of NGC 6621/2
corresponds to $M_B = -8.6$. As found for the Antennae by Whitmore
et al. (1999b), the contribution of the brightest stars is important
only fainter than $M_V = -8$ and negligible for $M_V > -9$, so even the bluest
of these will not be significant contaminants in our $B$ data. Likewise,
to these magnitudes, background galaxies are not major sources
of contamination.

We deal with two subsets of cluster candidates for each pair - 
a large sample selected for $ 4 \sigma$ significance in $B$, and
one culled for color accuracy, with  $\sigma (B-I) < 0.2$.
Photometric errors are based on photon statistics for aperture
photometry, and fitting errors for PSF-fitting measures.
The color samples have effective limits near $B=26$, somewhat
depending on local background intensity. For NGC 5752, we removed
about 40 faint cluster candidates (all fainter than $B=23.5$ and
most fainter than $B=25$) if their peak surface brightness was no
higher than found in an extended background region. This was to 
reduce false detections produced by clear patches between the
prominent dust lanes. 

\subsection{Fiber-array kinematic data}

To help understand the kinematics and interaction history of
NGC 6621/2, we measured the velocity field in H$\alpha$ and [N II]
emission using the 3.5m WIYN telescope and Dense-Pak fiber array.
As described by  Barden, Sawyer, \& Honeycutt (1998), the array
includes a $7 \times 13$ configuration of fibers in a roughly
hexagonal packing covering a $35 \times 45$" region, rotated for these 
observations to PA 114$^\circ$, to put the long axis approximately parallel 
to the galaxy separation in this pair. Four outlying fibers
allow sky subtraction far from the galaxy centers.
During 19/20 June 2000, four 25-minute exposures were
obtained in each of two regions, roughly centered on NGC 6621 and 6622.
Each of these sets was dithered in a parallelogram pattern, with
each leg offset by about 2", to fill the gaps between 3"
fiber apertures. The instrumental resolution with these fibers
was 1.6 \AA\ , well sampled by the 0.68-\AA\  pixel scale.
The spectral range observed was 6000-7400 \AA\ .
Velocity maps were produced from the H$\alpha$ and [N II]
$\lambda 6583$ measures; for each observation, emission was
detected in 43--57 fibers, for a total of 402 velocity measures. The
velocity maps were constructed on a 1" grid,
with overlapping aperture data averaged at each pixel and
numerical 0.1" subpixels used to track aperture outlines until the
final averaging. Registration to the direct images used reconstruction
of continuum and H$\alpha$ images from the spectral measurements, giving
positions of the galaxy nuclei and the bright cluster
of H II regions between the two.  

The resulting velocity field is shown in Fig. 4.
The long-slit velocity slice from Keel (1996) is also plotted for
comparison, preserving somewhat better spatial resolution
across the star-forming complex between the two galaxies.
The rotation curve is sufficiently asymmetric that this
pair (also known as CPG 534) was used in that paper as the type example
for how different ways of measuring redshifts from rotation
curve can differ for disturbed galaxies. In fact, we differ
in our conclusions with the study by 
Reshetnikov \& Sil'chenko (1990) as to the sense of the
encounter. They derive a low radial velocity for NGC 6622 and
conclude that the encounter is retrograde, noting that the
inner disk gas is rotating in the same sense as the outer
tidal features. We have no good explanation for this velocity discrepancy.
In retrospect, they may have been unlucky in
selection of spectroscopic position angles for analysis;
our velocity data are in reasonable agreement for regions
in common. Furthermore, the velocity field may be so
strongly disturbed as to make a set of tilted circular ring
orbits an inappropriate model in this system.

We use the WIYN spectra to address the redshift of NGC 6622. Because there 
is strong line emission around it which is associated with the distorted disk
of NGC 6621, and shows continuous velocity behavior across the
center of NGC 6622, it is unclear whether previously published
redshift values, especially from emission lines, actually reflect
the stellar component in NGC 6622. While the red region we
observed for gas kinematics is not ideal for absorption-line
redshifts, it includes the Na D lines at this redshift and
several weaker features such as the red Ca features near 6450 \AA\ .
Summing the four fiber spectra best centered on NGC 6622,
we can measure and resolve the Na lines (giving a Gaussian
velocity dispersion of about $\sigma_v= 315$ km s$^{-1}$). The
Na D equivalent widths, 1.7 and 1.6 \AA\ , are not much larger
than predicted from stellar population models, so that interstellar
absorption from foreground material does not dominate their measured
wavelengths. We also see weak emission in [N II] and filling of
expected H$\alpha$ absorption at the nucleus, close to the stellar
radial velocity. The mean
redshift from the sodium lines, $cz=6224 \pm 10$ km s$^{-1}$
in the heliocentric frame (within 1 km s$^{-1}$ of the
observed frame for this galaxy within $2.2^\circ$ of the ecliptic pole),
agrees well with the value $cz = 6241 \pm 10$ km s$^{-1}$
that we derive by cross-correlation of the NGC 6622 spectrum,
excluding the region near the Na D lines,
against a spectrum of M32 obtained with the same instrument
in December 2000. This is in contrast to a substantially lower
velocity of $5870 \pm 100$ km s$^{-1}$ reported by 
Reshetnikov \& Sil'chenko (1990). The sign of the relative 
velocity is important in understanding the pair's dynamics; our
measurement indicates a direct encounter, which fits with the
pair's overall morphology.

\subsection{Ground-based H$\alpha$ imaging}

To illustrate the surroundings of these pairs, and provide complementary
emission-line data on candidate clusters, we use ground-based $R$ 
and H$\alpha$ imagery obtained at the
prime focus of the 4m Mayall telescope at Kitt Peak National Observatory,
using a $2048 \times 2048$ Tektronix CCD. H$\alpha$ and [N II] emission
were isolated using a filter nominally centered at 
6709 \AA\ , shifted to about 6690 \AA\  in the converging $f/2.8$ beam, with
a FWHM of 71 \AA\  .
We used $R$-band images for continuum subtraction,
noting the requisite flux correction factor based on the ratio of
filter widths in deriving the H$\alpha$+[N II] fluxes. At 0.42"/pixel, the 
field includes wide areas around
each pair, which proved helpful in reducing problems from a reflection of
the telescope pupil by keeping the galaxies inside the central ``hole"
in this reflection. Surface-brightness calibration was carried out in $R$
using the NGC 7006 standard field as observed by Odewahn, Bryja, \& Humphreys 
(1992). The narrow-band image of NGC 6621/2 suffered from poor focus. The 
$R$-band image was convolved with a differential point-spread function (PSF)
to match the narrow-band image before continuum subtraction. This
still leaves significant residuals near the nucleus of NGC 6622,
as seen by comparison to the residuals near similarly bright stars.

Comparing the detections in H$\alpha$ and in the high-resolution HST continuum
images is important in constraining the ages of clusters, and in giving
hints as to how we should interpret a similar census of H II regions in
other galaxies from ground-based images. The surface-brightness sensitivity of 
the $R$ images is also helpful in finding faint outer tidal features, as in
Fig. 1. Minor-axis diffuse H$\alpha$
structure in NGC 5752 may indicate a starburst-driven wind.

The ionizing clusters are well enough separated in NGC 5754 that we
could derive H$\alpha$ equivalent widths for most of the HST
detections, fitting PSFs derived from star images and constrained to
lie at the locations measured with WFPC2. For H II regions that are
not strongly decentered from the starlight, this will give
improved accuracy in deblending partially overlapping images.
To do this, we resampled the KPNO images to match the coordinate systems of
WFPC2 mosaics, with object coordinates transformed from individual
CCDs to the same mosaicked pixel space. The IRAF implementation of
the DAOPHOT NSTAR routine was used to fit the fluxes of clusters in
both $R$ and H$\alpha$+[N II]. This process does not account for all
the H$\alpha$ emission; some is associated with clusters which are fainter 
than our error cutoff near $B=26$, some may be associated with optically
obscured ionizing sources, and some of the arm emission may be
truly diffuse. Cluster crowding in NGC 6621/2 means that we can
derive H$\alpha$ equivalent widths only for larger groupings of
clusters, and show evidence that some of the nuclear H$\alpha$
concentration arises in an outflow.
 
The $R$ images, with wide field and deep surface-brightness sensitivity,
are also useful in estimating luminosity ratios of the systems,
particularly NGC 5752/4. Here, we assume symmetry across the 
overlap regions, and derive an $R$-band intensity ratio of 5.1 for
the two galaxies, within an isophotal level corresponding to
a mean radius of 90".
 
\section{NGC 5752/4 - one with a starburst, one without}

\subsection{Dynamics and Interaction History}

The kinds of tidal structures seen in galaxy pairs can often show
important facets of their dynamical history. Some of the patterns seen in 
$n$-body simulations are robust to such parameters as halo/disk
ratio and disk thickness. In general, higher orbital inclinations
of the companion with respect to the target galaxy's disk plane
yield more asymmetric tidal structure, and direct encounters
produce larger-scale and more coherent tidal responses than do
retrograde encounters. We use the results from the simulation survey
by Howard et al. (1993) to deduce the kinds of encounter involved
in these two pairs, and estimate the time since strongest
perturbation. Their models incorporated 180,000 ``star" particles
and 20,000 ``gas" particles, distinguished by initial velocity
dispersion, and tracked their in-plane motions (while allowing the
point-mass companion to move and respond in three dimensions). 
Halo/disk ratios of 1 and 10 were considered, modelled as inert
Mestel disk distributions; the difference generally amounts to
having stronger fine structure in the low-halo case, as disk
self-gravity becomes more important.
This model matching will be less exact than attempts to simulate
a single system in detail, but we can exclude large parts of parameter
space and derive bounds on possible values of orbital inclination,
mass ratio, and impact parameter.

NGC 5754 shows a rich variety of structures which can be compared to
simulations, well shown in the wide-field depiction of Fig. 1 and
highlighted with our orbital reconstruction in Fig. 5. 
An inner two-armed grand-design spiral pattern twists around more
than $360^\circ$, with a single tidal narrow arm, not clearly extending the
inner pattern, wrapping $270^\circ$ around the side projected toward
the companion NGC 5752. One of the inner arms has a prominent kink
northeast of the core, where several luminous clusters and H II
regions occur. A fan-shaped feature (``spur") extends outward from one of
the arms on the northwest and northern sides of the inner disk, with
a more diffuse counterpart to the south and southeast.
NGC 5752 shows a long tidal tail stretching away from NGC 5754;
a similar feature stretching inward would be lost against the bright
disk of the larger galaxy. The WFPC2 image shows a sheet of dust in NGC 5752 
tracing a well-defined plane seen nearly edge-on, along the projected
major axis, which suggests that it, too, is a disk system.

We find the best overall fit to the structure of NGC 5754 from an encounter
inclined $60^\circ$ to its disk plane, with halo/disk mass ratio of 1,
from a low-mass companion passing just outside the disk. The mass ratio
derived this way, certainly between 0.1 and 0.5, is comparable to the
$R$-band flux ratio of the galaxies, 0.20. Fig. 5
illustrates the model structure two disk-edge rotations after the
initial close passage. This model reproduces the asymmetric (if
not quite one-sided) outer tidal arm, as well as the inner grand-design
pattern and fan-shaped disk structures. It also gives a change in
pitch angle for the northern arm at about the observed location.
The uncertainly in inclination is $\pm 15^\circ$, since there are
no good fits for $30^\circ$ or polar encounters. The mass ratio
could be 0.1-0.4, somewhat degenerate with perigalactic distance,
constrained to be from 1-2 radii of the initial disk (which itself corresponds
closely to the extent of the current grand-design pattern, based on comparison
of the model and image). The time scale can be set from the simulation,
given kinematic information on the system, since 
the orbital period at the edge of the initial
Mestel disk was set to $100 \pi$ timesteps. This radial location corresponds
to 36" (11.3 kpc) based on the structural match in Fig. 5. We derive
a circular velocity 285 km s$^{-1}$ from the H I profile parameters given by 
Haynes \& Giovanelli (1991), noting that their orientation estimate of
$25^\circ$ from face-on is close to the $27^\circ$ that we estimate from
comparison with the face-on model. These values give a rotation period at 
this radius of 220 Myr, and a timestep of 0.70 Myr. The best-fit model
is seen 350 timesteps after perigalacticon (which iccurs at time step
100 in the simulation shown in Fig. 5), so this pair
represents a relatively weak perturbation which
has developed through $\approx 2.5 \times 10^8$ years. The fit is
distinctly worse for models $\pm 50$ timesteps from this,
so 50 steps or 35 Myr is a reasonable error estimate.

Tidal distortions in the morphology and radial profile of NGC 5752
add clues to its orbital history. As discussed by Johnston (1998) and
references therein, tidal disruption of a small companion galaxy
results in stars becoming unbound and preferentially streaming
in tails oriented roughly along the orbit. Nearby examples include
the Magellanic Stream and the recently-identified tidal debris from
the globular cluster Pal 5 (Rockosi et al. 2002). As shown in Fig. 1,
NGC 5752 has a long, low-surface-brightness tail extending outward in
projection from NGC 5754; any such feature in the opposite direction 
would be lost against the much brighter structure in the larger galaxy's
disk. An isophotal twist occurs between the inner disk of NGC 5752
and this tail, accompanied by a break in the slope of the surface-brightness
profile and a change in the image ellipticity. These are shown in 
Fig. 6, from the results of using the STSDAS {\tt ellipse} task to fit
the KPNO $R$-band image, excluding the regions toward NGC 5754 where
spiral structure is bright and difficult to model easily.
Such a set of signatures
appears in models for the disruption of companion galaxies presented by
Johnston, Choi, \& Guhathakurtha (2002) and Choi, Johnston, \& Guhathakurtha
(2002). They show that the morphology of tidal debris depends 
on orbital eccentricity, phase, and viewing direction, and suggest ways
to distinguish whether the dominant mechanism for loss of stars is tidal
stripping (in near-circular orbits) or heating (for more eccentric paths).
If the break in the intensity profile occurs at the same radius as
the change in shape and orientation of the fitted ellipses
($r_{break} \approx r_{distort}$ in their terminology),
heating is dominant and the orbit has high eccentricity. For
near-circular orbits or near the apses of eccentric orbits, 
stripping dominates, so that isophotal twisting
which can be important well within the break radius 
($r_{break} > r_{distort}$). From Fig. 6, $r_{break} = 25$ arcsec in this
case, while distortion begins well inward of this, 
$r_{distort} \approx 16$". The photometric profile is thus consistent
, although not uniquely so, with our $n$-body interpretation, 
in which the initial path is parabolic
(before energy exchange at the first encounter). The direction of
the tidal debris also gives a hint at the orbital orientation, in the
sense that the rotation of position angles between the galaxy's disk
and tidal debris should go toward the direction of the orbit
(as shown, for example, by Johnston et al. 2002). Even the inner 
parts of NGC 5752 might be significantly prolate from tidal effects,
although the dust structure shows that disk symmetry is still dominant.
While the mass ratio in this pair makes the tidal impulse seen by
NGC 5754 relatively weak, NGC 5752 has undergone a much stronger
perturbation, although there is a smaller radial distance over which
it could operate.

\subsection{Star clusters and H II regions}

The WFPC2 data reveal a rich population of clusters in each galaxy, with very 
distinct properties. These offer an interesting comparison, especially
in regard to the different levels of tidal perturbation encountered by
the major and minor partners in the interaction, as well as clues to
when and where cluster formation can be triggered. An overview of these
populations is shown in Fig. 7, where the spatial distribution of clusters
is coded by their $B$ magnitudes.

\subsubsection{A starburst in NGC 5752}

NGC 5752 hosts numerous bright clusters, with a nearly flat differential
luminosity function from $B=21.6-26$. At the faint end, crowding is clearly an 
issue - the 
configuration is so small in projection that there is little room
for faint sources to be detected. The clusters span a broad color range 
$B-I=0.6-2.2$, with no particular evidence of color-magnitude
relation. Our criterion for useful color measurements of
$\sigma_{B-I} < 0.2$ retains 139 clusters out of
306 detected to $B=26$, with the errors dominated by background
noise from adjacent objects in the ``sky" annuli. Confusion between
genuine clusters and fine structure in the dust may set in, but
only fainter than about $M_B=-11$, an effect whose correction 
would accentuate the
true degree of weighting toward bright clusters that we observe.

A more important issue is the crowding effect of such a dense collection
of star clusters, in both object detection and photometry. We removed
candidates with overlapping photometric radii, retaining only the
brightest of each potential set. This leads to progressively more
severe undercounting of fainter clusters, as more of the available
area is already taken up by brighter clusters. We estimate the
effect of this crowding in two extremes. The minimum correction assumes
that the faint clusters are spread uniformly across the well-populated
part of NGC 5752 on the near side of the disk, about 27 square arcseconds.
The maximum correction assumes that they are concentrated to the central
region as strongly as the bright clusters, half of which occur
within a single region of 6.0 square arcseconds. The estimated
fraction of undercounting at a given flux is the ratio of total area
considered to the sum of the aperture areas for all brighter clusters.
For the faintest clusters with useful colors ($B \sim 24.8$), the
correction is a factor 1.4 for the minimum case, up to 3.0 for the
maximum case. Even the maximal correction leaves the luminosity function
unusually flat in comparison to most other systems.

The color-magnitude distributions and luminosity function of
these clusters are displayed in Fig. 8. Both stand in contrast to
the same diagrams for NGC 5754 (section 3.2). The distribution of
magnitudes is flatter, and extends brighter, in NGC 5752. The diagram
includes the luminosity function after applying the ``minimal" amd
``maximal" crowding corrections above. 

An interesting paradox is that the cluster population is also
redder. While NGC 5752 is visibly dusty, it is interesting that
only 5\% of this rich cluster population is as blue as the peak
in the color distribution of the ``normal" disk clusters in its neighbor
NGC 5754, which would require an extensive dust  distribution 
in addition to the obvious filaments.

The H$\alpha$ emission does not appear to be associated with individual
clusters, lacking peaks near bright clusters. Instead, there is
strong but diffuse emission following the disk of NGC 5752, plus
extended plumes along the minor axis (Fig. 9). More precisely stated,
any peaks associated with the brightest clusters are too weak to
stand out against the large-scale emission at 1" resolution. 
Comparison with the WFPC2 $I$ image blurred to the same resolution
shows that there may be a distinct peak corresponding to the two
bright clusters at the left end of the central region in Fig. 9,
but the match to the overall cluster distribution is not close.
The ridge line of H$\alpha$ is offset
northward from the bright central clusters, and the emission includes a 
significant component extending along the projected minor axis, most
prominent on the south side. Both these points are accounted for
by dust absorption in the disk plane, strnogly obscuring minor-axis structure 
on the far side). This appearance
is reminiscent of the filamentary emission from global winds in such
starburst systems as M82 and NGC 3628 (Shopbell \& Bland-Hawthorn 1998;
Fabbiano, Heckman, \& Keel 1990). Since such winds
may have ionization arising from shocks as well as from photoionization
by the galaxy's overall OB star population, H$\alpha$ will not necessarily give
a reliable measure of the ionizing luminosity of stars in this galaxy.
Virtually all the line emission is contained within a radius of
10", and indeed we see a monotonic decline in H$\alpha$+[N II]
equivalent width with increasing aperture radius, from 54 \AA\  at
5" to 28 at 12". These encompass the value of $43 \pm 5$ \AA\  listed by Kennicutt et al.
(1987). Both the star formation, as traced by bright clusters, and the 
ionized gas are
more concentrated than the red starlight, strongly enough that the
central H$\alpha$ equivalent width within 5"
indicates a burst of star formation unless most of the
line emission is powered mechanically rather than by photoionization.
Even in that case, such energy input must reflect a transient event,
so we have made a long-winded argument supporting what is clear from
the presence of any luminous clusters in a low-luminosity galaxy -
the interaction has triggered a burst of star formation. 

Some of the clusters' color spread is certainly due to dust reddening. Several of
the brightest clusters with $B-I > 2$ lie on the side of NGC 5752
where the dusty disk is in the foreground. In addition, the average
color of clusters in the inner 2" concentration is redder than
found in the surroundings, 1.9 versus 1.5. The full cluster population
is certainly richer than we see at these wavelengths, with additional 
examples hidden behind the dusty filaments in the disk plane.

\subsubsection{Normal disk clusters in NGC 5754}

In contrast, the clusters in NGC 5754 are fainter, and spread throughout
the disk and spiral pattern. Only one of these clusters would
appear among the fifty brightest (at $B$) in the smaller companion
system NGC 5752. The brightest clusters in NGC 5754 are in the
inner spiral pattern, just off the tips of the small bar. There are
no bright clusters inside the inner ring which surrounds the bar,
as is often found for H II regions in spirals with bars and rings
(Crocker, Baugus, \& Buta 1996).
The cluster population is well concentrated to the spiral arms,
a ``beads-on-a-string" trend which is strongest for the brighter clusters.
This is illustrated by overlaying the cluster positions 
on an $I$ image (Fig. 7). Of the clusters bright enough to
pass our $\sigma_{B-I} < 0.2$ criterion, 85\$ fall within the
continuum spiral arms. Looking at the entire cluster sample, interarm
clusters are as a group fainter than average, but faint enough that we
cannot tell whether they are also redder, as would be expected if
they are aging clusters which have moved outside the arm ridges where
they were formed. A handful of faint, very red clusters within the
ring may be genuine old globular clusters.

The color and magnitude statistics are summarized in Fig. 10, which
combines the color-magnitude diagram with histograms in $B$ magnitude
and color. All these distributions are quite distinct from the bright,
redder cluster population found in the companion galaxy NGC 5752.
The color span runs as blue as the bluest unreddened value
$B-I = -0.3$ expected for a cluster formed instantaneously, with
most clusters bluer than $B-I = 0.7$. This suggests that most of
these clusters are younger than $7 \times 10^7$ years, based on the
Starburst99 models. The luminosity function in this galaxy is
strongly weighted to faint clusters, again in contrast to the flatter
distribution in NGC 5752, where our maximum crowding correction 
gives $\alpha = -1.7$.

The luminosity function of clusters can often be expressed in a simple
power-law form. As noted by Larsen (2002), a power-law function in
luminosity $N(L) dL = \beta L^\alpha$ implies linear behavior in
number per unit magnitude $N(M) = b + a M$ where $\alpha = - (2.5 a + 1)$. 
For clusters brighter than $B=25$, a least-squares fit to the
binned luminosity function (for bins with $n > 1$, using 
Poisson weights) has $a=0.69$, yielding $\alpha=-2.7 \pm 0.4$. 
This is not unusual for a ``normal" spiral disk; Larsen (2002)
finds undisturbed spirals with $\alpha = -2.0-2.5$. Larsen also
shows that a constant star-formation rate gives a similar value
$\alpha = -2.7$. However, for cluster age ranges comparable to
the dynamical timescales of many interactions, the slope of the
cluster luminosity function is fairly constant; Whitmore et al. (1999b)
find that the bright clusters in NGC 4038/9 exhibit $\alpha = -2.6 \pm 0.2$
in $V$, with younger clusters showing some evidence of a flatter
slope (as in their Fig. 18). In contrast, for the range of crowding corrections
used for NGC 5752, its cluster system has $\alpha > -1.7$.

The combination of age and mass effects has made interpreting the
luminosity function alone of clusters (or H II regions) ambiguous. 
An important role for age is suggested by the
reddening of mean colors for fainter objects in NGC 5754
(as in Fig. 10). However, potential selection effects make it
difficult to make such a case from the WFPC2 data alone, since extreme
colors will drive up the $B-I$ error and drop clusters from the
sample, and systematic behavior of reddening with location across
spiral arms could mimic age behavior as well.
Multiple continuum colors can improve the
situation, as can emission-line information.
We can further examine the history of cluster formation
in NGC 5754 by comparing continuum colors 
with H$\alpha$ equivalent widths, using a hybrid data set of HST aperture 
photometry and KPNO H$\alpha$ data. The H$\alpha$ fluxes
and associated $R$ continuum fluxes were measured by PSF fitting 
on the KPNO images, at
cluster locations fixed from the HST identifications. As shown
in an overlay of the H$\alpha$ image over the WFPC2 B data
(Fig 11), much of the line emission is associated with identified
clusters. By the same token, most of the bright clusters have
H$\alpha$ counterparts, which immediately shows that they are bright
because of youth rather than extreme mass. The
line emission reflects young, massive stars, type B0 and hotter,
so its strength changes dramatically with age over timespans
that leave $B-I$ nearly constant. The comparison is thus
sensitive to the ages of clusters and the distribution of
ages in the sample. These
data (Fig. 12) show an envelope of decreasing H$\alpha$ EW for
redder clusters, indicating that age is a more important factor
in the color spread of this cluster population than is reddening. 
If these clusters are single-burst systems formed over less than a
few million years, the youngest we see is about $8 \times 10^6$ yr old, 
with many at ages up to about $1.5 \times 10^7$ yr. If each cluster we see
incorporates stars formed over a span of a few $10^6$ yr, the derived ages
are smaller, since the $B-I$ color will reflect significant contributions
from the first stars to form red supergiants while the H$\alpha$ equivalent
width is dominated by younger, hotter members. Durations of a few million
years for each star-forming event (cluster) would account for 
the maximum observed H$\alpha$ equivalent width near 200 \AA\ 
and the distribution of observed colors and equivalent width for the
cluster population.

\subsection{History of Cluster Formation}

The NGC 5752/4 system provides a striking contrast between disks which
underwent dramatically different levels of tidal perturbation at the
same time. Nothing very unusual seems to happened in NGC 5754; its cluster
population is comparable to other similarly luminous spirals such as NGC 6946
and NGC 4258 from Larsen (2002). In contrast, NGC 5752 has hosted a
starburst with a rich population of very luminous clusters.
This is consistent with the idea that there is a minimum level of
tidal influence needed to trigger the formation of these massive clusters.

The clusters in NGC 5752 are
unusually red and have a wide color range, which fits with the observed
dust structure in suggesting substantial reddening. None of the clusters
in NGC 5752 is as blue as the
population in NGC 5754 over the same magnitude ranges. 
The least reddening that we can assign without multicolor
data assumes the clusters to be concentrated in age so as to
populate the reddest excursion in predicted colors, around $1.2 \times 10^7$
years. Most clusters would have to fall in this range, however, and
about one third of them still require foreground dust. Combining
dust and age fading, this population of clusters must have been
even more spectacular a few times $10^7$ years ago. 

In this instance, much of the cluster formation must have been confined
to this short span. The evidence for an outflow raises the possibility that
cluster formation has been terminated by loss of dense gas, which
might be related to the flat (or even peaked) luminosity function of these
clusters. Such a distribution is particularly interesting as a possible
step toward the narrow Gaussian luminosity function of
globular clusters. By analogy with formation of stars in 
such regions as the Eagle Nebula, we might speculate that lower-mass
clusters would suffer first as a wind develops.

\section{NGC 6621/2 - rapid response to a strong perturbation}

\subsection{Dynamics and Interaction History}

The major tidal feature in NGC 6621 is the extensive one-sided
tidal arm wrapping to its north. Additional starlight and H II regions
span the region between the galaxy nuclei, and continue past the
core of NGC 6622. A dust lane in front of NGC 6622 shows a helical
twist, which indicates that it is close enough in three dimensions to
be strongly accelerated by the smaller galaxy's potential. 
As in the case of NGC 1409/10 (Keel 2001), such twists may
be a signature of pairs favorable for mass transfer between members.
NGC 6622 shows a narrow edge-on disklike structure spanning the inner 7"
and a dust feature spanning about 1.5" (Fig. 3), inclined by about 30$^\circ$ 
to the apparent stellar disk and not obviously connected to the extensive 
dust in NGC 6621. This is the only disturbed feature of NGC 6622.
The emission-line velocity field (Fig. 4), and new redshift measure for
NGC 6622, confirm the directions of relative motion as being a direct
encounter for NGC 6621 and nearly polar one for NGC 6622.
 
The extensive tidal disruption in NGC 6621/2 immediately suggests
a mass ratio near unity, and a small impact parameter. Indeed, we find the
best morphological match for equal-mass galaxies, closest approach
near the disk edge and much less than two disk radii, unit halo/disk
ratio, and an early time. We are then seeing this system within
half an initial disk-edge rotation period of closest approach,
corresponding to a time since perigalacticon of about 
$1.0 \times 10^8$ years using the inner symmetric rotation curve
in Fig. 4.
As shown in Fig. 13, this model reproduces the extensive splash of
material between the galaxies, the strongly curved tidal arm on the
opposite side, and even the presence (though not the distribution) of
material diverging from a close approach to 
NGC 6622. The behavior of 
such material depends strongly on the perigalacticon distance,
mass distribution and precise location of
NGC 6622. In this pair, NGC 6621 is undergoing a strong perturbation,
and we view it shortly after closest approach and maximum tidal stress.
Since our target pairs were selected from tidal morphology,
it is no coincidence that the best match in each case is for a
direct encounter.

Seeing NGC 6621/2 early in a strong encounter may allow us a close
look at where the clusters form, since there has been little time
for clusters triggered by the interaction to diffuse very far from their
formation sites, either as considered within the rotating disk of
NGC 6621 or in the context of the overall tidal structure.

\subsection{Cluster Population}

NGC 6621 has a very rich population of luminous star clusters. They are
found near its nucleus, in the inner disk largely toward NGC 6622,
in a prominent star-forming complex between the two nuclei, and in the
opposing tidal arm (Fig. 14). Their color and magnitude distributions
are shown in Fig. 15. Clusters extend as bright as $M_B=-14.6$ and
as blue as the youngest modelled populations from the Starburst99
code. The color distribution peaks near $B-I = 0.8$, bluer than we
see in the clusters of NGC 5752. The circled points in this plot
show clusters in the complex between the nuclei, discussed in the
following section.

As in the case of NGC 5754, we can fit a power law to the luminosity function
at bright levels where incompleteness is not an issue. Here we derive
the logarithmic slope of the magnitude distribution as $a=0.50 \pm 0.05$
implying a power-law index in luminosity 
$\alpha = - 2.25 \pm 0.12$, again in the range found for both
normal and starburst galaxies.

There is an asymmetry in number of clusters between ``inner" and ``outer" halves of
NGC 6621, seen in both the HST continuum results and the H$\alpha$
structure (Fig. 16). This fits generally with tidal triggering of
cluster formation, since the tidal stress has been much stronger
in the region between nuclei, where we see more clusters. As a result, 
perturbations of the velocity field
have been stronger (as shown in Fig. 4), which must logically precede
disturbances in gas density or cloud separation.

Clusters, with and without strong H$\alpha$ emission, are seen
in the prominent tidal arms or tail to the northeast. Cloud
collapse is still going on in this region, as has
been noted in NGC 4676 by de Grijs et al. (2002). Indeed,
the existence of H II regions in the tidal tails of NGC 4676
and 4038/9 has long shown that star formation can continue for
several times $10^8$ years in gas launched well away from 
interacting systems into tidal features. This was reflected in 
the relatively strong H$\alpha$ emission measured in the tails of
the Mice by Stockton (1974) and in the Antennae by Schweizer (1978)
and Mirabel et al. (1992).
At its extreme, such processes may be symptoms of the existence
of massive and self-bound clumps in the tails, such as would
form tidal dwarf galaxies (although the dynamics of their formation
may require a role for dark matter, as discussed by Hibbard et al. 2001).

The broad color distribution indicates that many of the clusters in
NGC 6621 are significantly reddened and dimmed. From the reddest
excursion of the colors in the Starburst99 models, clusters are
present with at least $A_{B-I}=0.7$, and as large as 2.0 if
other models with a less pronounced redward loop are accurate.
This makes the cluster population intrinsically even more
impressive, especially since the second brightest cluster
at $B$ must be reddened by at least $E_{B-I}=0.9$ which means
that its corrected absolute magnitude at least as bright as
$M_B=-16.3$. Several additional clusters would likewise
stand only about 0.5 magnitude fainter with this minimum
reddening correction. Cluster masses must range up to at least
$6 \times 10^5$ solar masses using the Starburst99 tracks.
The bluest clusters give a minimum age, requiring that they
be less that $10^7$ years old.

\subsection{A Massive Disk Star-forming Region}

One of the most striking features of NGC 6621/2 is the very luminous
star-forming region found along the line between the nuclei, 
at a projected distance of 26" or 11 kpc from the center of NGC 6621.
A long-slit spectrum (Keel 1996) shows interesting kinematics
in this region, with the slope of the rotation curve reversing
in sign for about 2 kpc. The WFPC2 images break this region
up into over 40 individual clusters, which are among the bluest
to be found in the whole system. They are marked as circled data
points in Fig. 15. Most of the disk H$\alpha$ emission comes from
this area as well, also marking this as a current site of extraordinary
star formation. The projected dimensions of this region are
$ 2.6 \times 5.5$" or $1.1 \times 2.3$ kpc, with the direction of elongation 
indicating
that it  is likely to be as narrow as it looks but quite possible longer;
simply projection in the plane at an inclination of $70^\circ$ would
make it $1.1 \times 6$ kpc, so that seen face-on this might appear
a very bright piece of the spiral structure.

The overall H$\alpha$ equivalent width in this region is the highest
found in the entire system, at 285 \AA\  averaged on 2" scales. This
corresponds to ages $6-7 \times 10^6$ years in the Starburst99 models.
Star formation in this area is a recent phenomenon in comparison to the
interaction timescale. Significant older populations would decrease
the H$\alpha$ equivalent with, and the observed distribution in
$B-I$ does not allow any of the detected clusters to be older
than $10^8$ years even neglecting reddening.
Obvious dust clouds near the ``intergalactic" complex
suggest that extinction may be important in some of these
cluster colors, though it would have to exceed $A_B=1$ to hide
luminous clusters from detection. Thus some of the cluster colors
are likely to be even bluer than we measure.

\section{Analysis}

\subsection{Formation mechanisms for massive clusters}

Multiple mechanisms have been suggested as drivers
for the enhanced star formation during galaxy encounters. 
The locations of star formation, especially in massive clusters,
during the encounters we have observed may help refine
the list, since some kinds of dynamical processes will operate 
preferentially in certain parts of the galaxies or timespans.
Our own results can be supplemented with those for other well-observed
and reasonably well-modelled interacting systems. The best-known is NGC 4038/9,
one of the pairs whose striking tidal features inspired the original
Toomre \& Toomre (1972) investigation using restricted 3-body calculations. 
Their rough orbital
history has been substantially refined by successive $n$-body models of
increasing sophistication and accordingly
better matches to the observed velocity field as well as morphology.
We adopt the orbital parameters used by Barnes (1988) and Mihos et al. (1993).
This puts our present view about $2.1 \times 10^8$ years after the
initial close passage between the galaxies, whose equal-mass disks had
radii about 10 kpc and with a smallest nuclear separation of 13 kpc.
They are now well into the second approach of a rapidly decaying mutual
orbit, with the next merging passage to come in about $10^8$ years.
Both disks see a passage inclined by about $60^\circ$. Since this
situation is more complex than the single dominant close passage we deal 
with in the other systems, and the systems remain nearly in disk contact,
we take this system as representing the results of an interaction seen
in mid-encounter. 

For NGC 2207/IC 2163, we use the model described by Elmegreen et al. (2000)
with refinements provided by C. Struck. These galaxies have approached
on a highly eccentric relative path, with their disks grazing
about $4 \times 10^7$ years before our current view.
The relative path is inclined by $25^\circ$ to the disk of NGC 2207,
which is the more massive galaxy by a factor close to 3. The properties of
the four systems we compare are given in Table 1, including statistics
of their brightest star clusters.

Collisions between molecular clouds are an obvious mechanism for
driving up the star-formation rate during galaxy encounters, and
offer an attractive way to compress the large amounts of gas needed to
make SSCs. We can rule out a necessary role for collisions between
clouds from different disks by the fact that we see a rich population of
SSCs in NGC 6621 whose companion is apparently an S0 system and
of small radius by comparison, and in NGC 5752, which has remained well
beyond the gas-rich regions of NGC 5754 during the encounter. In fact,
each of these galaxies has more of the brightest clusters ($M_B < -12$) than
NGC 4038/9.

Cloud collisions for clouds originating in the
{\it same} disk should be tracked for an in-plane encounter by the locations 
of orbit crossing for the clouds.
However, the loci of clusters in NGC 6621/2 do not
match the expected sites of orbit crossings. For example, the simulation
by Klari\'c \& Byrd (1990) shows the orbital streamlines for a similar
encounter, in which crossing occurs predominantly on the leading
edge of the bridge structure between the nuclei. The Howard et al. (1993)
results give a similar picture, though in less detail because velocity
information was not saved and can be only partially derived from cloud
locations in successive snapshots. The star formation, as seen from
both the cluster locations and H$\alpha$ mapping, is concentrated
near the center of the bridge region, between the nuclei, and in the
opposite tidal arm. Similarly, the few luminous clusters in IC 2163
are not found in the tidal arm, shown by the Elmegreen et al. (2000) 
model to have rapid crossing of disk orbits from material initially
in different parts of the disk. (We note that some pairs do show
extensive star formation in roughly the regions expected for orbit crossing;
a good example is NGC 6745).

Another dynamical mechanism may be at work here, suggested by the
flattening (and indeed local reversal) of the rotation curve near the
luminous star-forming complex. Toomre (1964) presented a stability criterion 
for the velocity dispersion of gas in rotationally supported disks, 
$$ \sigma_v > 3.36 G \mu / \kappa \eqno{(1)},$$
where $G$ is the gravitational constant, $\mu$ is the disk surface density
at the relevant point, and $\kappa$ is the epicyclic frequency for perturbations
about a circular orbit, which may be expressed in terms of the
angular velocity $\Omega (R)$ as
$$ \kappa = [R {{d \Omega^2}\over {dR}} + 4 \Omega^2]^{1/2}\eqno(2)$$
(e.g. Binney \& Tremaine 1987).
This criterion is violated within radii interestingly close to 
the outermost star formation in many disks (Kennicutt 1989), which
has been interpreted as meaning that this instability influences the
overall dynamics of the interstellar medium. In a strong interaction,
as seen in NGC 6621/2, the local velocity structure can be dramatically
changed, and a large enough ripple in rotation velocity (and
associated effective potential) can drop the epicyclic frequency
$\kappa$ over regions large enough to allow substantial gas masses
to go unstable. Numerical simulations (e.g. Klari\'c 1993) show that
this instability is strong enough to allow a wide range of
linear scales to become unstable on comparable timescales.
This may also help explain the very large gas concentrations found
by McCain (1997) in spiral/elliptical pairs, in which enough gaseous
mass accumulates to alter the local disk kinematics (see also Bransford 
et al. 1999). The result may appear in NGC 6621 as a complex
like the ``interface" cloud in NGC 4038/9, but with less extinction.
This is the most active star-forming site in the system, and the uniformly
blue cluster colors suggest that it may have begun only recently.

It is noteworthy that the SSC populations in NGC 6621 and NGC 2207 
concentrate on one side of the disks, the one toward the companion.
The question of whether star-forming regions in interacting galaxies
favor the (current) direction of the companion goes back at least
to a debate between Arp (1973) and Hodge (1975) on the brightest H II
complexes in disturbed galaxies. Arp's sample suggested that they
occur preferentially between the two nuclei, while Hodge's analysis
indicated that they fell opposite the companion nucleus as often
as between the two. More detailed study has been hampered by projection
effects; we do not always know which side of each galaxy as we see it
is closer to the other one in space. Even crude matches to numerical simulations
can be useful in telling which pairs we can understand well enough to
recognize inner and outer sides in. If indeed SSCs are seen preferentially 
between the
galaxies, this says that the formation trigger operates quickly and
gravitationally, since it must track the current location of the
other galaxy. For penetrating encounters such as NGC 4038/9, 
cloud collisions can act in this way, but for more distant encounters
as in NGC 6621/2 and NGC 2207/IC 2163, a less direct process must be at work.

\subsection{Timing and Duration of Cluster Formation}

Most well-studied SSC populations are in currently merging systems and aging merger 
remnants.
Our data bear on the processes happening before mergers, and in
more distant encounters which happen well before the final merging
approach. The color behavior of SSCs in the older, post-merger systems shows
a general match with the dynamical ages of these objects, so that the
cluster formation is roughly coeval with merging. Looking at galaxies
in earlier stages can tell more precisely when, and over what interval, 
cluster formation is triggered.

The bluest cluster colors in each of these galaxies show that cluster
formation is continuing, extending so blue as to
require cluster formation within the last $10^7$ years. This is
less than the time since strongest tidal perturbation for each
interaction (although the components of NGC 4038/9 are so close that
it is fair to characterize this interaction as ``ongoing"). Cluster
formation continues for a long time after the perturbation begins and
peaks, and may be delayed. 

Barton et al. (2000) have found that the integrated rate of star formation
in galaxy pairs, as traced from H$\alpha$ emission, correlates with
projected linear separation and velocity difference so as to suggest
that there is a delay of a few times $10^7$ years between perigalactic passage
and the onset of enhanced star formation, which lasts $\sim 10^8$ years.
Such a delay makes sense for any sort of kinematic trigger, since
the velocity perturbation which responds immediately to changes
in the potential much precede morphological disturbance, including
changes in the distribution both stars and gas.
This is consistent with the patterns we see in luminous clusters, a tracer
which is detectable longer than the H$\alpha$ emission. In some regions,
such as the young group of clusters and H II regions in NGC 6621/2,
the star formation must have set in very recently, as the cluster color
distribution and overall H$\alpha$ equivalent width do not allow a long
history of star formation. The association of H II regions with
so many luminous clusters shows that active star formation continues
well after the closest approach.

The redder cluster population in NGC 5752, and the evidence for a global
wind seen in H$\alpha$ emission, raise the question of whether its
starburst has been recently quenched as the wind removes gas from the
inner regions. We certainly expect this to happen in mergers, and 
extension to single disturbed galaxies could be relevant to 
stellar population in I0 galaxies and the E+A or K+A galaxies seen
in a variety of environments.

\section{Summary}

We have studied the star-cluster populations of galaxies in the interacting 
pairs NGC 5752/4 and NGC 6621/2 using WFPC2 images and dynamical 
models. These pairs were selected to include a range of interaction
age and strength of tidal perturbation, so as to help isolate the
parameters that are most important in triggering the formation
of massive super star clusters (SSCs). NGC 5752/4 is seen about 
$2.5 \times 10^8$
years after closest approach in a system with a mass ratio near 0.2,
while NGC 6621/2 is a system about $10^8$ years after a near-grazing
encounter of roughly equal-mass galaxies. Their star-formation responses
are correspondingly varied. NGC 5754 saw a mild, if prolonged, perturbation, 
and has a normal population of disk clusters. Its companion, which
saw a proportionally stronger tidal impulse, has a rich and red
population of SSCs, concentrated to the center and with a flat
luminosity function. Cluster formation may have been interrupted
by a starburst wind seen in H$\alpha$. NGC 6621 has a very rich
population of SSCs, with young clusters concentrated in a small region
between the galaxy nuclei. 

This would fit
with the frequent location of the brightest star-forming in pairs
between the two nuclei, suggesting a nearly instantaneous trigger,
and favor star formation in different locations than a cloud-collision
trigger.

We have compared the SSC populations in these pairs to those of two
additional pairs undergoing current interactions, and for which
dynamical modelling gives a fairly accurate timescale, NGC 2207/IC 2163
and the Antennae, NGC 4038/9. This comparison demonstrates several
points which are important in understanding the formation of massive
star clusters. Rich cluster populations do not require direct contact
between the stellar disks of the galaxies, and seem to require
some minimum level of tidal disturbance regardless of the time
allowed for its impact to develop within a disk. Furthermore, the cluster
distributions do not match predictions for the locations of orbit
crossing within a single perturbed disk. The major
collection of SSCs and H II regions in NGC 6621 lies in an area
where the slope of the velocity field reverses, leading us to 
explore the idea that
gravitational disturbance of a disk may allow gas to become instable
by the Toomre criterion over previous stable regions, a process which
could act quickly mediated solely by the perturber's gravitational
influence. This would fit
with the frequent location of the brightest star-forming in pairs
between the two nuclei, suggesting a nearly instantaneous trigger,
and favor star formation in different locations than a cloud-collision
trigger.

In some regions, extensive star formation has begun only
recently compared to the interaction timescale, consistent with
evidence for a delayed onset of enhanced star formation in interacting
galaxies found from global measures. In one case, the cluster colors
and H$\alpha$ morphology suggest that a starburst wind has quenched the
formation of new clusters. This object, NGC 5752, also shows a flatter
cluster luminosity function, which might be related
to the difference between power-law forms seen for young clusters
in most galaxies and the near-Gaussian form for old globular clusters.

The HST archive is now rich in imagery suitable for surveying the brightest
clusters in interacting pairs, and prospects are promising for improved
understanding through additional studies involving dynamical modelling
and supporting ground-based H$\alpha$ data.

\acknowledgments

This work was supported by NASA through STScI grants GO-07467.01-96A,B.
Thanks to Di Harmer and Gillian Rosenstein for helping make
efficient use of a 4-hour break in the clouds while working at the WIYN
telescope. Martha Holmes helped with the imaging at Kitt Peak, as part of
an NSF Research Experiences for Undergraduates program.
We thank Brad Whitmore for providing data on NGC 4-038/9 in advance of 
publication, and Gene Byrd for conversations on the art of reconstructing
interactions. We acknowledge the public service provided by Claus Leitherer
and coworkers in providing the Starburst99 code and its results
for general use. Likewise, we acknowledge the value of the $z$-machine
data archive recently made available by the Center for Astrophysics.
Curt Struck provided his most recent model parameters for the NGC 2207/IC 2163
in advance of publication. Eija Laurikainen first mentioned the potential role
of the Toomre instability to us several years ago. We thank the referee for
useful comments, particularly in evaluating the role of crowding in
NGC 5752.

\clearpage
\figcaption
{The two galaxy pairs of Arp 297, marked on a KPNO 4m $R$-band image.
Velocities are heliocentric. The WFPC2 observation footprint is indicated
around NGC 5752/4. The image is shown with a pseudologarithmic intensity 
mapping to emphasize faint tidal structures.
The field spans $5.16 \times 6.19$ arcmin.
\label{fig1}}

\figcaption
{WFPC2 imagery of NGC 5752/4, shown as a color composite incorporating
a synthetic green band taken as the mean of $B$ and $I$, with a
pseudologarithmic intensity scale. The regions shown covers
$90 \times 108$"; north is approximately $18^\circ$ counterclockwise from the
top. The cluster population in the large spiral NGC 5754 (left) stands out in 
color as well as brightness. In the companion galaxy NGC 5752 (right), the 
clusters are brighter and more
neutral (redder) in color; reddening in the dust lanes as well
as from more diffuse material on the south side of the galaxy is
well shown.
\label{fig2}}

\figcaption
{WFPC2 imagery of NGC 6621/2, again shown in a color composite incorporating
a synthetic green band taken as the mean of $B$ and $I$, with a
pseudologarithmic intensity scale. The region shown is $64 \times 122$",
and north is $46^\circ$ clockwise from the top. A rich population
of bright blue clusters and highly structured dust absorption are
clearly shown. A small patch of dust nearly in front of the nucleus of
NGC 6622 (left) is nearly lost in this display.
\label{fig3}}

\figcaption
{Radial-velocity data for the NGC 6621/2 system.
The upper panel overlays the WIYN fiber-array velocity field on
the WFPC2 $I$ mosaic. The dashed curves outline the regions with
detected line emission. Labelling of the contours is intended to be
sufficient to show the direction of slope in each region. The
sampling apertures were 3" in diameter, setting the spatial resolution 
of the combined velocity map.
The lower panel shows a velocity slice across the midline of the system 
(along the thin line in the image) from
long-slit data (Keel 1996, with $\pm 2 \sigma$ error bars), with somewhat 
higher spatial resolution,
showing the velocity structure across the powerful star-forming region
between the two nuclei. The data sets are aligned so that corresponding
points match along the $x$-axis. The nucleus of NGC 6622 has an absorption-line 
radial velocity $cz = 6241 \pm 10$ km s$^{-1}$, putting it close to the 
leftmost point of the longslit data. This is important in confirming that
this encounter is in a direct sense, as indicated by numerical
simulations like that shown in Fig. 13.
\label{fig4}}

\figcaption
{Simulation and our orbital reconstruction of the encounter of NGC 5752/4.
The $R$-band image (left, as in Fig. 1) has been annotated to show
the important tidal structures we attempt to match from the simulation.
The best-matching numerical simulation is shown at the right to the same scale
and orientation (found for an inclination $26^\circ$ from face-on).
The relative orbit of NGC 5752 is shown approximately, with the
grey curve indicating portions behind the disk of NGC 5754; uncertainties
in the system's inclination make this less accurate then the match of
structures in NGC 5754. The sequence at the bottom shows the simulation
from which the best match was drawn, at intervals of 50 timesteps or 35
million years; the enlargement on the right shows the fourth of these at $t=450$.
Perigalacticon occurs at $t=100$ timesteps, so our view is 250 Myr later. 
The features matched include the presence of an inner
grand-design spiral, the strong asymmetry in intensity between the two
outer tidal arms, and the location of the two spiral ``spur" features and
the northern kink in the main spiral pattern. The simulation frames
represent ``gas" particles, which show the characteristic structures
more clearly than the ``star" particles and will represent young stellar
populations more closely. The asymmetric tidal arms are also present at
later times, but the spiral ``spurs" fade at later timesteps.
\label{fig5}}

\figcaption
{Radial profiles of the $R$-band structure from NGC 5752, derived from ellipse 
fitting to the western half of the galaxy to reduce contamination from
NGC 5754. Following Fig. 3 of  Johnston et al. 2002, we indicate
distortion and profile-break radii, whose ratio suggests that the
relative orbit of NGC 5752 has large eccentricity. To emphasize the
break in surface-brightness profile, the two dotted lines indicate
exponentials fit for $r = 2-18"$ and $r > 27"$, with an outer
scale length almost four times larger than the inner value.
\label{fig6}}

\figcaption
{The locations of star clusters in the NGC 5752/4 system marked on the $I$
WFPC2 mosaic. These comprise the ``bright" sample, cluster candidates with
color error $\sigma _{B-I} < 0.2$. Except in the crowded regions of
NGC 5752, this corresponds approximately to $B < 26$. The orientation is
as in Fig. 2, enlarged to show a region $77 \times 104$" in extent. 
The arm kink used in matching the numerical model is marked, along with
the spur extending out of the frame.
\label{fig7}}

\figcaption
{Color-magnitude diagram for clusters in NGC 5752, including those
with errors $\sigma_{B-I} < 0.2$. The wide color spread is independent
of magnitude, and the differential luminosity function (shown in
number of clusters per 0.5-magnitude bin) for detected 
clusters is nearly flat from $B=21.5-24.5$; the dotted histogram
includes clusters which do not satisfy the color-error criterion for the
main panel, which are most detections at $B>24.5$. The two uppermost
histograms show this distribution after applying our minimum and maximum 
crowding corrections.
The two brightest clusters
from NGC 5754 are shown as open circles for comparison with Fig. 10.
The effective magnitude limit for this sample is probably no fainter
than $B=24.5$, limited by crowding in the inner part of the galaxy.
The color evolution of a single cluster (scaled to $10^5$ solar masses)
is shown from the Starburst99 output described by Leitherer et al. 
(1999), as is a reddening vector for this filter system and a young
stellar population. The age of the Starburst99 models is shown by crossmarks 
every $5 \times 10^6$
years up to $10^8$; the first of these is almost lost in the initial
loop near $B-I=0$. The reddest color in the model occurs at $8.6 \times 10^6$
years as red supergiants reach their peak contribution. Part of this red
loop (from 5--10 Myr)is repeated for referece at a stellar mass of $10^4$ 
solar masses,
showing that the masses of the detected clusters mostly lie in the range
$10^4 - 3 \times 10^5$ solar masses.
The right-hand panel shows the color distribution of all clusters plotted in 
the color-magnitude array.
\label{fig8}}

\figcaption
{H$\alpha$+[N II] emission from NGC 5752, shown as contours
overlaid on the F450W WFPC2 image. The region shown is 20 arcseconds
square, with the contours spaced logarithmically at intervals of 0.2 dex
and the image also logarithmically mapped. The orientation is as in
Figs. 2 and 7, with north 18$\circ$ counterclowkwise from the top.
The WFPC2 image as shown has been mosaicked, losing some resolution, because 
the PC frame includes no objects suitable for tying together the
astrometric frames of WFPC2 and the H$\alpha$ image. 
The minor-axis plume suggests that
much of the H$\alpha$ emission is associated with a global wind,
while the poor match to the cluster distribution indicates that very little 
of the emission comes from H II regions around individual ionizing clusters.
\label{fig9}}

\figcaption
{Color-magnitude diagram for clusters in NGC 5754, including those
with errors $\sigma_{B-I} < 0.2$ and laid out as in Fig. 8. This population 
is simultaneously bluer
and fainter than that seen in NGC 5752. The cluster masses must
be substantially smaller as well, an effect which will be
amplified by the allowed range of reddening in the smaller
companion.
The dashed luminosity function includes clusters
with color errors larger than $\sigma_{B-I}=0.2$, giving a sample 
complete to a level perhaps 0.5 magnitude deeper than the color sample.
\label{fig10}}

\figcaption
{H$\alpha$ contours from the continuum-subtracted KPNO image, overlaid
on the WFPC2 $B$ image of NGC 5754. The $B$ image is displayed logarithmically, 
while the contours in this case are spaced linearly. This region is 50" across.
The comparison illustrates the level of correspondence
between HST cluster detections and H II regions, which is good but not
completely one-to-one. A good example comes from the clump of clusters 
to the upper right of the nucleus, where some have strong H$\alpha$ and
others show no influence on the H$\alpha$ structure.
\label{fig11}}

\figcaption
{Color-emission strength relation for clusters in NGC 5754, with H$\alpha$
derived by PSF fitting on the KPNO images with positions fixed to
match the cluster locations from WFPC2 data. For a foreground screen,
a reddening vector is horizontal, while for mixed stellar
and dust distributions, the situation becomes complex. The
plotted curve is the trajectory for an instantaneous burst
of star formation from the Starburst99 code, at times from $5-15 \times 10^6$
years (top to bottom).
\label{fig12}}

\figcaption
{Comparison of tidal structures in NGC 6621/2 with a numerical model,
laid out as in Fig. 5. The relative orbit is strongly foreshortened in this 
case, with closest approach occurring almost in front of the center of NGC 6621.
A perigalacticon slightly farther out rapidly suppresses the extent of
the tidal ``spray" of particles which pass close to the core of NGC 6622,
leading us to assign a closest approach somewhat greater than the
precomputed model shown here. Closest passage occurs at about
$t=175$, which puts the observed configuration about 
75 timesteps or $1.0 \times 10^8$ years later.
\label{fig13}}

\figcaption
{Locations of cluster candidates with color error $\sigma_{B-I}<0.2$ in the
NGC 6621/2 system, oriented as in Fig. 3.  In most parts of the system, this
error bound corresponds roughly to $B < 26$. This region is
$40 \times 112$" in extent.
\label{fig14}}

\figcaption
{Color-magnitude array for clusters in NGC 6621/2, arranged as in 
Figs. 8 and 10.  The peak in color near
$B-I = 0.75$ gives limited scope for reddening of the bright 
clusters and requires that they be younger than about $10^8$ years
from continuum colors alone.  However, the blue edge to
the cluster distribution suggests that the overall rate of
cluster formation is now declining, and some clusters must
be reddened by effective amounts as large as
$A_B=2$. Circled points indicate clusters in the crowded
star-forming complex between the nuclei, which are systematically both
the brightest and bluest in the entire system. The cluster luminosity
function shows evidence of flattening to faint magnitudes at a level
not seen in NGC 5754, despite similar overall crowding. The Starburst99
model track is the same as in Figs. 8 and 10. 
\label{fig15}}

\figcaption
{H$\alpha$ contours overlaid on the WFPC2 $I$ image of NGC 6621/2. 
The contours are logarithmic in emission-line intensity, at flux
intervals of $\surd 2$. The central peak is significantly offset from
the
nucleus of NGC 6621, along the minor axis, suggesting that much of this
H$\alpha$ emission comes from an outflow associated with the nuclear
star formation. The greatest H$\alpha$ equivalent width, by nearly a factor 2,
is found in the star-forming complex between the two nuclei. 
\label{fig16}}

\clearpage

\begin{table*}
\begin{center}
\begin{tabular}{lccccc}
\tableline
\tableline
Object          &          NGC 5754 & NGC 5752  &  NGC 6621 &   NGC 2207  &  NGC 4038\cr
Orbital inclination &        60$^\circ$ &  60$^\circ$: &  0$^\circ$ &   25$^\circ$ & 60$^\circ$ \cr
Closest passage (disk radii)       &    1.5 &   2: &  1.2     &  1.0    & 0.8 \cr
Time since then (Myr)              &   250  &  250 &   100    &   40    &   0 \cr
Mass ratio                         &   0.2  &  5   &   1      &   0.3   &    1\cr
Brightest cluster ($M_B$)          & -12.8  & -13  &   -14.5  &  -14.1  &  -14\cr
Bluest SSC,  $B-I$                 &  -0.1  & -0.1 &   0.05   &   -0.3  &  -0.12\cr
Clusters with $M_B<-12$ &   1    &  20  &   200    &    2    &   18\cr
\tableline
\end{tabular}
\tablenum{1}
\caption{
Interaction and cluster properties \label{tbl1}}
\end{center}
\end{table*}

\end{document}